\let\oldvec\vec
\documentclass[runningheads]{llncs}

\let\vec\oldvec

\usepackage{times}
\usepackage{tikz}
\usepackage{amssymb}
\usepackage{amsmath}
\usepackage{complexity}
\usepackage{enumitem}
\usepackage{graphicx}

\newtheorem{mytheorem}{Theorem}
\newtheorem{myobservation}{Observation}
\newtheorem{mydefinition}{Definition}
\newtheorem{mylemma}{Lemma}
\newtheorem{mycorollary}{Corollary}
\newtheorem{myremark}{Remark}
\newtheorem{myproposition}{Proposition}

\newcommand{\myqed}{\mbox{$\Box$}}

\begin{document}

\title{Almost Envy Freeness and Welfare Efficiency \\ in Fair Division with Goods or Bads}
\titlerunning{Almost Envy Freeness and Welfare Efficiency in Fair Division with Goods or Bads}

\author{Martin Aleksandrov}
\authorrunning{M. Aleksandrov}

\institute{TU Berlin, Germany}

\maketitle              

\begin{abstract}
We consider two models of fair division with indivisible items: one for goods and one for bads. For goods, we study two \emph{generalized} envy freeness proxies (EF1 and EFX for goods) and three common welfare (utilitarian, egalitarian and Nash) efficiency notions. For bads, we study two \emph{generalized} envy freeness proxies (1EF and XEF for goods) and two less common \emph{dis}welfare (egalitarian and Nash) efficiency notions. Some existing algorithms for goods do \emph{not} work for bads. We thus propose several \emph{new} algorithms for the model with bads. Our new algorithms exhibit many nice properties. For example, with additive identical valuations, an allocation that maximizes the egalitarian diswelfare or Nash diswelfare is XEF and PE. Finally, we also give simple and tractable cases when these envy freeness proxies and welfare efficiency are attainable in combination (e.g.\ $0/1$ valuations, $0/-1$ valuations, house allocations).

\end{abstract}

\section{Introduction}

Fair division is the task to allocate a number of items to a number of agents \cite{brams1996,brandt2016,moulin2003}. If the items are \emph{goods} (e.g.\ cakes, real estates, food), agents would tend to act selfishly and thus receive more items. If the items are \emph{bads} (e.g.\ house-hold chores, project tasks, job shifts), agents would tend to cooperate and thus share more items. Many fair division models consider items as goods, e.g.\ rent divisions \cite{gal2018}, credit assignments \cite{clippelmoulin2008}, cake divisions \cite{aziz2016bef}, land allocations \cite{halevi2015}, memory distributions \cite{parkes2015}, rooms scheduling \cite{bouveret2017,smet2016}, etc. Some other fair division models consider items as bads, e.g.\ tasks assignments \cite{heydrich2015,verne2016}, chores divisions \cite{aziz2017}, etc. Yet another fair division models suppose a mixed manna of goods and bads \cite{aziz2018ai3,bogomolnaia2016gab,bogomolnaia2016gob}. In this paper, we consider two fair division models: one with goods and one with bads. We thus study axiomatic and computational connections between these models. 

The ``golden'' axiomatic standard of fair division is envy freeness (i.e.\ no agent envies another agent) \cite{foley1967}. Envy freeness (EF) may unfortunately not be attainable (e.g.\ one item and two agents). Many approximations of EF for goods have thus been proposed \cite{amanatidis2018,aziz2018,lan2010}. Two such proxies are envy freeness up to \emph{some} good (EF1) and envy freeness up to \emph{any} good (EFX). That is, by eliminating (some or any) single good, we can attain EF. EF1 is always attainable while it is unknown if EFX is always attainable \cite{caragiannis2016}. Similar proxies were recently adapted for goods and bads \cite{aziz2018ai3}. The ``golden'' axiomatic standard in economics is efficiency (e.g.\ Pareto efficiency (PE) or welfare efficiency (WE)) \cite{clippel2008,varian1974}. Efficiency may always be attained as opposed to EF. 

In our work, we propose two proxies of EF for the model with bads (i.e.\ 1EF and XEF). We thus study axiomatic and computational connections between 1EF and XEF for bads and EF1 and EFX for goods. We first argue that our notions are more general than existing notions as they require the \emph{marginal valuation} of an agent for an item to be positive. Many existing notions require the \emph{valuation} of an agent for an item to be positive \cite{amanatidis2018,budish2010,caragiannis2016}. As a consequence, these existing notions are trivially satisfied in problems in which each agent values positively only bundles of two, three or more items. Our notions are moreover robust to \emph{zero} marginal valuations as well. Some existing notions allow the valuation of an agent for an item to be zero. However, in this case, it may be \emph{impossible} to achieve envy freeness up to any good (i.e.\ EFX) \cite{plaut2018}. 

We also analyze these axiomatic properties in combination with welfare efficiency (i.e.\ utilitarian, egalitarian or Nash). For example, with identical valuations, an allocation in a problem with goods is EFX (or EF1) and WE iff it is XEF (1EF) and WE in the \emph{negated} problem with bads (i.e.\ all valuations are negated). In contrast, this relation does \emph{not} hold with additive distinct valuations and even 2 agents. There is a simple reason why we focus on WE and PE. Welfare efficiency is more demanding and challenging than Pareto efficiency. An allocation that is welfare efficient is also Pareto efficient whereas the opposite does not hold. Moreover, we can achieve ``almost'' envy freeness and welfare efficiency in combination. For example, with additive valuations for goods, an allocation that is Nash efficient is also EF1 \cite{caragiannis2016}. For bads, questions in a similar vein remain unanswered \cite{aziz2018ai3}. In response, we show that an allocation that maximizes the \emph{Nash diswelfare} (i.e.\ the product of disutilities) is only guaranteed to be PE.

We are also interested in whether algorithms for ``almost'' fair division problems with goods can be applied to ``almost'' fair division problems with bads. For example, the \emph{Lipton algorithm} returns an allocation of goods that is EF1 even with monotone valuations \cite{lipton2004}. However, Lipton's allocations may not be welfare efficient \cite{plaut2018}. By comparison, in \cite{caragiannis2016}, the authors show that the exponential \emph{MNW algorithm} is Nash efficient and EF1 with additive valuations. With 0/1 valuations, the \emph{Alg-Binary algorithm} returns an EFX and welfare efficient allocation in polynomial time \cite{barman2018}. Even more, the \emph{leximin++} solution is EFX and PE with monotone identical valuations and the \emph{Alg-Identical algorithm} returns a allocation that is EFX and welfare efficient with additive identical valuations \cite{barman2018,plaut2018}. However, none of these algorithms can be directly applied to a given problem with bads.   

For this reason, most of these algorithms do \emph{not} return fair allocations with bads. We thus propose a number of new algorithms for the setting with bads, i.e.\ \emph{Lipton-- algorithm}, \emph{leximax-- solution} and \emph{Alg-Identical--}. For example, unlike \emph{Lipton algorithm}, our \emph{Lipton-- algorithm} is 1EF with anti-monotone valuations. Further, unlike the \emph{leximin++} solution, our \emph{leximax--} solution is an XEF allocation. We also show that our \emph{Alg-Identical-- algorithm} returns the same allocation in a given problem as the \emph{Alg-Identical algorithm} in the negated problem. We, hence, conclude that this allocation is XEF and welfare efficient. We thus view the \emph{Alg-Identical-- algorithm} as an alternative to the \emph{Alg-Identical algorithm}. Moreover,we identify a simple case (0/-1 valuations) when XEF and welfare efficiency are attainable. Finally, we prove that it is \emph{not} possible to achieve 1EF and welfare efficiency with anti-monotone identical valuations.

\section{Related work}

We only discuss the work that is most related to our work. EF1 was initially proposed for a model with ordinal preferences for goods \cite{budish2010}. Instead, we consider cardinal valuations for goods or bads. EFX for goods was proposed in \cite{caragiannis2016}. Some variants of these notions are considered in \cite{amanatidis2018,aziz2018ai3,barman2018}. As discussed earlier, these notions are trivially satisfied in problems in which no agent has positive value for any single item. We thus believe that our notions are \emph{more general} than existing notions. Other notions of ``almost'' envy freeness allow zero marginal valuations \cite{plaut2018}. Instead, our notions do not. We believe that thus our notions are \emph{more robust} to small perturbations of the problem input than existing notions. In \cite{aziz2018ai3}, the Lipton's algorithm is generalized to work with additive valuations for bads (and goods). In contrast, we generalize this algorithm to work in fair division problems in which agents have arbitrary anti-monotone valuations for bads. Moreover, our algorithm bounds the maximum envy in such problems.

From a welfare perspective, various objectives have been considered for fair division with bads, e.g.\ the product of disvaluations, the product of valuations \cite{bogomolnaia2016gab,bogomolnaia2016gob}. The product of valuations is sensitive to small changes in the problem input, e.g.\ agent leaves or agent arrives. The product of disvaluations 
may not provide any fairness guarantees as well \cite{aziz2018ai3}. We show that minimizing this product is tractable and guerantees Pareto efficiency. Moreover, this product is monotone in each agent's valuation and it satisfies scale invariance (if an agent doubles all her valuations this does not change which outcomes maximize the objective) \cite{freeman2017}. By comparison, we also study the minimum disvaluation in an allocation. Interestingly, maximizing it is XEF and welfare efficient with additive identical valuations or PE with anti-monotone identical valuations.

We further view our work as an overview of existing algorithms for fair division with goods (e.g.\ the \emph{leximin++} solution) \cite{barman2018,caragiannis2016,lipton2004,plaut2018}. However, we also propose \emph{new} algorithms for fair division with bads (e.g.\ the \emph{leximax--} solution). These algorithms are in some sense computationally equivalent to the existing algorithms. For example, the \emph{leximin++} solution is hard even to approximate \cite{chakrabartychuzhoy2009,dobzinski2013,goemans2009}. This is also true for our \emph{leximax--} solution. In fair division with goods, ``almost'' envy freeness and Nash efficiency cannot be achieved with monotone identical valuations \cite{caragiannis2016}. In fair division with bads, we also give similar impossibility results. Finally, fair division algorithms for goods have already been ``unleashed'' in practice \cite{goldman2014}. We hope that our work motivates the development of such algorithms for bads.

\section{Preliminaries}

We consider a set $N=\lbrace 1,\ldots,n\rbrace$ of agents and a set $O=\lbrace o_1,\ldots,o_m\rbrace$ of indivisible items. We suppose that each $i\in N$ has a \emph{valuation} function $u_i: 2^O\rightarrow\mathbb{R}$ that assigns $u_i(B)$ to each $B\subseteq O$. An \emph{instance} is the triple $\mathcal{I}=(N,O,(u_i)_{i=1}^n)$. If the items are \emph{goods}, we suppose that $u_i(B)\in\mathbb{R}_{\geq 0}$. If the items are \emph{bads}, we suppose that $u_i(B)\in\mathbb{R}_{\leq 0}$. We consider monotone valuations for goods \emph{and} anti-monotone valuations for bads. We say that $u_i(B)$ is \emph{monotone} iff, for each bundle $C\subseteq O$ with $C\supseteq B$, $u_i(C)\geq u_i(B)$. We say that $u_i(B)$ is \emph{anti-monotone} iff, for each bundle $C\subseteq O$ with $C\supseteq B$, $u_i(C)\leq u_i(B)$. We suppose that $u_i(\emptyset)$ is zero. We further write $u_i(o)$ for $u_i(\lbrace o\rbrace)$. A valuation $u_i(B)$ is \emph{additive} if it is $\sum_{o\in B} u_i(o)$.

We write $A=(A_1,\ldots,A_n)$ for an \emph{allocation} of the items from $O$ to the agents from $N$ where (1) $A_i$ is the bundle of items of agent $i$ for each $i\in N$, (2) $\cup_{i\in N}^n A_i=O$ and (3) $A_i\cap A_j=\emptyset$ holds for each two agents $i,j\in N$ with $i\not=j$. We say that an allocation $A$ is \emph{envy free} (EF) iff, for each $i,j\in N$, $u_i(A_i)\geq u_i(A_j)$. Also, an allocation $A$ is \emph{Pareto efficient} (PE) iff, there is no other allocation $A^{\prime}$, $u_i(A^{\prime}_i)\geq u_i(A_i)$ for each $i\in N$, and $u_k(A^{\prime}_k)> u_k(A_k)$ for some $k\in N$. An allocation $A$ is \emph{max welfare efficient} wrt welfare $w$ (MW) iff $w$ is maximum under $A$. An allocation $A$ is \emph{min welfare efficient} wrt welfare $w$ (mW) iff $w$ is minimum under $A$. For goods, we consider the \emph{utilitarian welfare} (UW) $\sum_{i\in N} u_i(A)$, \emph{egalitarian welfare} (EW) $\min_{i\in N} u_i(A)$ and \emph{Nash welfare} (NW) $\prod_{i\in N} u_i(A)$. For bads, we also consider the \emph{egalitarian diswelfare} (EDW) $\min_{i\in N} (-u_i(A))$ and \emph{Nash diswelfare} (NDW) $\prod_{i\in N} (-u_i(A))$.

\subsubsection{Envy freeness proxies} Two EF proxies that have been proposed for fair division with goods are envy freeness up to \emph{some} good and envy freeness up to \emph{any} good \cite{caragiannis2016}. We identify a problem with existing definitions of these two properties and their variants \cite{aziz2018ai3,barman2018,plaut2018}. With monotone valuations, these existing properties are trivially satisfied in problems in which no agent has a positive but zero valuation for an item. We, therefore, propose \emph{new} and more general definitions of these proxies for goods.

\begin{mydefinition} $($\emph{EF1 allocation}$)$
An allocation $A$ is \emph{envy free up to some good} (EF1) iff, for $i,j\in N$ with $A_j\not=\emptyset$, $\exists o\in A_j:u_i(A_j)-u_i(A_j\setminus\lbrace o\rbrace)>0$, $u_i(A_i)\geq u_i(A_j\setminus\lbrace o\rbrace)$. 
\end{mydefinition}

\begin{mydefinition} $($\emph{EFX allocation}$)$
An allocation $A$ is \emph{envy free up to any good} (EFX) iff, for $i,j\in N$ with $A_j\not=\emptyset$, $\forall o\in A_j:u_i(A_j)-u_i(A_j\setminus\lbrace o\rbrace)>0$, $u_i(A_i)\geq u_i(A_j\setminus\lbrace o\rbrace)$.    
\end{mydefinition}

With goods, an EF allocation is further an EFX allocation that is even further an EF1 allocation. With bads, EFX and EF1 are trivially satisfied as no agent has positive marginal valuation after the removal of an item from another agent's bundle. For this purpose, we propose two alternative proxies of EF in this setting.

\begin{mydefinition} $($\emph{1EF allocation}$)$
An allocation $A$ is \emph{envy free up to some bad} (1EF) iff, for $i,j\in N$ with $A_i\not=\emptyset$, $\exists o\in A_i:u_i(A_i)-u_i(A_i\setminus\lbrace o\rbrace)<0$, $u_i(A_i\setminus\lbrace o\rbrace)\geq u_i(A_j)$. 
\end{mydefinition}

\begin{mydefinition} $($\emph{XEF allocation}$)$
An allocation $A$ is \emph{envy free up to any bad} (XEF) iff, for $i,j\in N$ with $A_i\not=\emptyset$, $\forall o\in A_i:u_i(A_i)-u_i(A_i\setminus\lbrace o\rbrace)<0$, $u_i(A_i\setminus\lbrace o\rbrace)\geq u_i(A_j)$.    
\end{mydefinition}

With bads, an EF allocation is an XEF allocation that is even an 1EF allocation. Interestingly, proxies for goods and proxies for bads relate but only in the special case with \emph{identical} valuations (i.e.\ each agent has the same valuation for each bundle).

\begin{myremark}\label{rem:one}
With identical valuations, an allocation in a problem with goods is EFX (EF1) iff it is XEF (1EF) in the negated problem with bads.
\end{myremark}

\begin{myremark}\label{rem:two}
Even with 2 agents and additive distinct valuations, an EFX (EF1) allocation in a problem with goods may not be an XEF (1EF) allocation in the negated problem with bads, and an XEF (1EF) allocation in a problem with bads may not be an EFX (EF1) allocation in the negated problem with goods.
\end{myremark}

EF1, EFX, 1EF and XEF coincide with existing properties for additive valuations. Moreover, our definitions are robust to zero ``one-item'' marginal valuations. Finally, we use throughout the paper EF, EF1 and EFX in the setting with goods, and EF, 1EF and XEF in the setting with bads. 

\section{Algorithms}

In our work, we use a number of existing algorithms. These algorithms were originally proposed for fair division with goods.

\begin{itemize}[topsep=0pt]
 \item \emph{Lipton algorithm}: with monotone valuations, for each good, the algorithm bounds the maximum envy of the current allocation, thus obtaining a new allocation and allocating the good to an agent such that nobody \emph{envies} them \cite{lipton2004}.
 \item \emph{MNW algorithm}: with additive valuations, (1) it computes a \emph{largest} subset of agents with positive valuations for goods, and (2) it then computes an MNW allocation over this subset \cite{caragiannis2016}.
 \item \emph{leximin++ solution}: with monotone valuations, it first maximizes the \emph{minimum} valuation in an allocation and \emph{maximizes} the size of the bundle of an agent with such minimum valuation, followed by maximizing the second \emph{minimum} valuation and then \emph{maximizing} the size of the second minimum valuation bundle, and so on \cite{plaut2018}.
 \item \emph{Alg-Identical algorithm}: with additive identical valuations, it orders the goods by \emph{decreasing} valuations and, for each next good, it allocates the good lexicographically to an agent with \emph{minimum} total valuation for the previous goods \cite{barman2018}.
 \item \emph{Alg-Binary algorithm}: with additive 0/1 valuations, see \cite{barman2018} for details.
\end{itemize}

We show that some existing algorithms do not work in fair division with bads. In response, we propose a number of \emph{new} algorithms in this setting.

\begin{itemize}[topsep=0pt]
\item \emph{responsive draft algorithm} agents pick their most preferred remaining items in some strict priority ordering, and they miss their turns if they value positively no remaining good.
 \item \emph{Lipton-- algorithm}: with anti-monotone valuations, for each bad, the algorithm bounds the maximum envy of the current allocation as the \emph{Lipton algotihm}, thus obtaining a new allocation and allocating the bad to an \emph{envy free} agent. 
 \item \emph{leximax-- solution} (a portmanteau of ``lexicographic'' and ``maximax''): with anti-monotone valuations, it first \emph{minimizes} the \emph{maximum} valuation in an allocation and \emph{minimizes} the size of the bundle of an agent with such maximum valuation, followed by \emph{minimizing} the second \emph{maximum} valuation and then \emph{minimizing} the size of the second maximum valuation bundle, and so on.
   \item \emph{Alg-Identical-- algorithm}: with additive identical valuations, it orders the bads by \emph{increasing} valuations and, for each next bad, it allocates the bad lexicographically to an agent with \emph{maximum} total valuation for the previous goods.
\end{itemize}

\section{Fair division with goods}

We consider the \emph{problem} of fair division with monotone valuations for goods. We are particularly interested in which envy free proxies can be guaranteed to exist and can be achieved simultaneously with various notions of efficiency.

\subsection{EF1 and EFX} With monotone valuations, an EF1 allocation is shown to exist and can be computed in $O(mn^3)$ time by using the \emph{Lipton algorithm} \cite{lipton2004}. With additive valuations, this can be done in $O(m)$ time by using a \emph{responsive draft algorithm} \cite{caragiannis2016}. In a similar vibe, an EFX allocation is shown to exist \emph{but} only when limited to monotone identical valuations, or 2 agents with monotone but possibly distinct valuations \cite{plaut2018}. However, with 3 or more agents and even additive distinct valuations, the problem if such allocations exist remains \emph{open} for several years. We leave it open. Instead, with 0/1 additive valuations, we show that an EFX allocation always exists and can be returned by a \emph{responsive draft algorithm} in linear time. 

\begin{myproposition}\label{pro:one}
With 0/1 additive valuations, an EFX allocation can be computed in $O(m)$ time.
\end{myproposition}

\subsection{EF1, EFX and Efficiency} We first consider additive valuations. For distinct valuations, even with 2 agents, an MUW allocation may give to one of the agents most of the most valued goods of the other agent. Such an allocation is PE but it cannot be fair. At the same time, an MEW allocation is EF1 (even EFX) but only with 2 agents  (see Example C.2 in \cite{caragiannis2016} and Theorem 5.5 in \cite{plaut2018}). Additionally, in this setting even with 3 goods, we can show that \emph{no} allocation can be MEW and MNW at the same time. For these reasons, we next focus on the egalitarian welfare and Nash welfare in isolation. We start with the egalitarian welfare. With identical valuations, the \emph{leximin++ solution} is guaranteed to be EFX and PE (Theorem 5.4 in \cite{plaut2018}). Unfortunately, even approximating it is intractable \cite{chakrabartychuzhoy2009,dobzinski2013,goemans2009}. A naive brute-force algorithm for it could take $O(2^m)$ time at worst. This might be too slow given that there are only $O(mn)$ valuations unless $m$ is a constant.

\begin{myobservation}\label{obs:one} $($\emph{Theorem 5.4 in \cite{plaut2018}}$)$
With $m$ goods additive identical valuations, an EFX, PE, MUW and MEW allocation can be computed in $O(2^m)$ time.
\end{myobservation}

We continue with the Nash welfare. With possibly distinct valuations, an MNW allocation is EF1 and PE for any number of agents (see Theorem 3.2 in \cite{caragiannis2016}). Interestingly, such an allocation is even EFX and MUW with identical valuations. In this case, MUW follows trivially because each allocation has the same utilitarian welfare.

\begin{mytheorem}\label{thm:one}
With additive identical valuations, an MNW allocation is EFX, PE and MUW.
\end{mytheorem}

\begin{myproof}
Let $A=(A_1,\ldots,A_n)$ be an MNW allocation. Hence, it is EF1 and PE. MUW follows with identical valuations. We next show that $A$ is even EFX. We consider two cases. In the first case, there are $m\leq n$ goods. The allocation $A$ is EFX because each agent receives in it at most one good. In the second case, there are $m>n$ goods. The Nash welfare of $A$ is positive in this case. For the sake of contradiction, assume that $A$ is not EFX. WLOG, we can suppose that agent 1 is not EFX of agent 2. Therefore, $u_1(A_1)<u_1(A_2\setminus\lbrace o\rbrace)$ for some $o\in A_2$. We next consider the allocation $B=(B_1,\ldots,B_n)$ with $B_1=A_1\cup\lbrace o\rbrace$, $B_2=A_2\setminus\lbrace o\rbrace$ and $B_i=A_i$ for each $i\in\lbrace 3,\ldots,n\rbrace$. We next show that the Nash welfare of $B$ is strictly greater than the Nash welfare of $A$ and thus reach a contradiction with the MNW of $A$. To do this, consider the ratio between the welfare of $B$ and the welfare of $A$. We derive that the value of this ratio is greater than one iff agent 1 is not EFX of agent 2. The result follows.
\myqed
\end{myproof}

We can compute an MNW allocation by using the \emph{MNW algorithm}. A fast implementation of this algorithm is presented in \cite{caragiannis2016}. However, we can show that its worst-case running time is $O(S(m,n))$ ($S(m,n)\approx\frac{n^m}{n!}$ is the Stirling number of second kind). Indeed, with $m>n$ goods, we cannot even hope for a pseudopolynomial-time algorithm that computes such an allocation \cite{nguyen2013}. Surprisingly, with $m\leq n$ goods and even distinct valuations, we can do this in $O(n^3)$ time.

\begin{mytheorem}\label{thm:two}
With $m\leq n$ goods and additive valuations, an EFX, PE and MNW allocation can be computed in $O(n^3)$ time.
\end{mytheorem}

\begin{myproof}
Let us add $(n-m)$ ``dummy'' goods to $O$ that are valued with 0 by all agents. Construct a weighted, undirected and complete bipartite graph $(N,O,w)$ such that $w((i,o))=\ln u_i(o)$ if $u_i(o)>0$ and $w((i,o))=0$ otherwise. A perfect matching in this graph is a set of $n$ vertex-disjoint edges. Let us consider the allocation $A$ that corresponds to a maximum weight perfect matching in this graph. The allocation $A$ is EFX because each agent receives in it exactly one good. Further, $A$ is over a largest set $N_A$ of agents to which one can simultaneously provide a positive valuation for a good because the sum of the valuations in it is maximized. Moreover, the Nash welfare over agents from $N_A$ is maximized because the sum of the logarithmic valuations over these agents is maximized. If $m=n$ and $|N_A|=n$, then the Nash welfare of $A$ is positive and maximized. If $m<n$ or $m=n$ and $|N_A|<n$, then the Nash welfare of $A$ or any other allocation is equal to 0. In both cases, $A$ is PE because the Nash welfare over agents from $N_A$ and the Nash welfare over agents from $N$ are both maximized. Finally, we can compute a maximum weight perfect matching in the graph in $O(n^3)$ time by using the Hungarian method \cite{kuhn1955}.\myqed
\end{myproof}

We might be satisfied with an ``almost'' MNW allocation in case it takes too long to compute such an ``exact'' allocation . For example, with $m>n$ goods and identical valuations, an EFX allocation might be ``almost'' MNW up to a tight factor of $1.061$ (see Lemma 4.2 and Example 4.3 in \cite{barman2018}). That is, the Nash welfare of such an allocation is between $\frac{1}{1.061}$ multiplied by the maximum Nash welfare and the maximum Nash welfare. With $m\leq n$ goods and such valuations, an EFX allocation can be both ``exact'' MNW and MEW. In both settings, we can return such allocations with the deterministic greedy \emph{Alg-Identical algorithm} (see Lemma 4.1 in \cite{barman2018}). By comparison, the \emph{MNW algorithm} runs in $O(n!)$ time to return such an ``exact'' allocation. 

\begin{myobservation}\label{obs:two} $($\emph{Theorem 3.1 in \cite{barman2018}}$)$
With $m>n$ goods and additive identical valuations, an EFX, PE, MUW and ``almost'' MNW allocation can be computed in $O(m)$ time.
\end{myobservation}

\begin{mytheorem}\label{thm:three}
With $m\leq n$ goods and additive identical valuations, an EFX, PE, MUW, MEW and MNW allocation can be computed in $O(m)$ time.
\end{mytheorem}

\begin{myproof}
We observe that an allocation in this setting is MNW iff it is MEW. This holds because the valuations are identical and an optimal allocation is over a largest set of agents to each of which one can simultaneously provide a positive valuation for a good. By definition, in this setting with $m\leq n$ goods, the \emph{Alg-Identical algorithm} maximizes the number of agents receiving exactly one good. Consequently, it returns an allocation that is EFX, PE, MUW, MEW and MNW in $O(n)$ time.
\myqed
\end{myproof}

By Observations~\ref{obs:one},~\ref{obs:two} and Theorem~\ref{thm:three}, we conclude that there is a computational trade-off between MEW and ``almost'' MNW with $m>n$ goods. This trade-off vanishes with $m\leq n$ goods and identical valuations. In the general case, we could try to satisfy only EF1 and PE in the absence of both MEW and MNW. For example, with 2 agents, an allocation that satisfies both EF1 and PE can be computed in linear time (see Theorem 2 from \cite{aziz2018ai3}). With 3 or more agents and identical valuations, we can also do this in linear time by Observation~\ref{obs:two} and Theorem~\ref{thm:three}. With distinct valuations, polynomial algorithms such as the \emph{responsive draft algorithm} and \emph{Lipton algorithm} cannot guarantee PE \cite{caragiannis2016,plaut2018}. In fact, we believe that \emph{no} other polynomial algorithm can guarantee EF1 and PE in isolation without MEW or MNW (unless $\P=\NP$). There might be problems with complex combinatorial valuations in each EF1 and PE is also an MEW or MNW allocation. Thus, in such problems, \emph{no} algorithm would run fast. To illustrate our idea, consider an easy problem in which there are 2 agents with values $2,1$ and $1,2$ for 2 goods. Each allocation that is not MEW or MNW either violates EF1 or PE.

Finally, when limited to simple 0/1 valuations, an EFX allocation can be optimal for any type of efficiency. We can compute such an allocation in $O(2m(n+1)\cdot\ln(mn))$ time with the deterministic greedy \emph{Alg-Binary algorithm} \cite{barman2018}. The allocation returned by this algorithm is MNW. Therefore, it is EF1. In this setting, an allocation is EF1 iff it is EFX. Moreover, this allocation gives to each agent only items they value positively. Consequently, it is PE and MUW. And, we can show that it satisfies MEW.

\begin{myproposition}\label{pro:two}
With 0/1 additive valuations, the \emph{Alg-Binary algorithm} returns an EFX, PE, MUW, MEW and MNW allocation in $O(2m(n+1)\cdot\ln(mn))$ time.
\end{myproposition}

We next consider monotone valuations. With \emph{zero} marginal identical valuations, there are problems in which \emph{no} EF1 allocation may be PE even with 2 agents (see Theorem 3.3 in \cite{caragiannis2016} and see Theorem 5.2 in \cite{plaut2018}). With \emph{non-zero} marginal identical valuations, EFX and PE may not be attainable again even with 2 agents (see Theorem 5.6 in \cite{plaut2018}). In contrast to this result and Theorem~\ref{thm:one}, we show that \emph{no} EF1 allocation may be MNW (a subset of PE allocations).

\begin{mytheorem}\label{thm:four}
Even with 2 agents and (non-zero marginal) monotone identical valuations, there are problems in which \emph{no} MNW allocation is EF1.
\end{mytheorem}

\begin{myproof}
Let us consider $N=\lbrace 1,2\rbrace$ and $O=\lbrace o_1,o_2,o_3\rbrace$. Further, let $u(O)=8$, $u(\lbrace o_1,o_2\rbrace)=7$, $u(\lbrace o_2,o_3\rbrace)=u(\lbrace o_1,o_3\rbrace)=5$, $u(o_1)=u(o_2)=4$ and $u(o_3)=3$. Consider the allocation $A=(\lbrace o_1,o_2\rbrace,\lbrace o_3\rbrace)$. This allocation achieves Nash welfare with value of 21. Any other allocation achieves Nash welfare with some lower value. Hence, $A$ is Nash efficient. However, it is not envy free up to some good (i.e.\ EF1) as $u(A_1\setminus\lbrace o_2\rbrace)=4>3=u(A_2)$ and $u(A_1\setminus\lbrace o_1\rbrace)=4>3=u(A_2)$ hold.\myqed
\end{myproof}

By comparison, the \emph{leximin++ solution} is EFX (and EF1), PE and MEW (see Theorems 4.2 and 5.4 in \cite{plaut2018}). We give a simple ``cut-and-choose'' protocol for this solution. This protocol is based on the idea that an agent can ``cut'' the set of goods into $n$ bundles that are mutually EFX from their point of view by simply computing the \emph{leximin++ solution} with all $n$ agents as they have identical valuations. Therefore, these bundles are mutually EFX from each other agent's perspective. Thus, each agent can now ``choose'' a different bundle in some strict order. Finally, with distinct valuations, EFX and MEW are attainable only with 2 agents (see Theorems 4.3 and Figure 5 in \cite{plaut2018}). 

\section{Fair division with bads}

We consider the \emph{problem} of fair division with anti-monotone valuations for bads. The \emph{reduced} problem is the sub-problem restricted to the bads valued negatively by all agents. The \emph{negated} problem is the problem in which all valuations are negated. As for goods, we are interested in which proxies exist and can be combined with efficiency.

\subsection{1EF and XEF} We start with 1EF. For additive valuations, an 1EF allocation always exists and can be computed in $O(m)$ time by using a \emph{responsive draft algorithm}. For anti-monotone valuations, we can apply the \emph{Lipton algorithm} directly to the problem. Unfortunately, the algorithm allocates all bads to a single agent. This is not fair to them. Instead, we can run the \emph{Lipton algorithm} on the negated problem and return an allocation that is EF1 in it. By Remarks~\ref{rem:one} and~\ref{rem:two}, this allocation is further 1EF in the initial problem but only with identical valuations. We, therefore, conclude that we need a \emph{new} algorithm for the general case. Our \emph{Lipton-- algorithm} is such an algorithm.

\begin{mytheorem}\label{thm:five}
With anti-monotone valuations, an allocation with bounded maximum envy can be computed in $O(mn^3)$ time.
\end{mytheorem}

\begin{myproof}
Let $\alpha=\max (u_i(B)-u_i(B\cup\lbrace o_j\rbrace))$ be the maximum marginal valuation over $B\subseteq O$, $i\in N$ and $o_j\in O$. Further, in a given $A$, let $e_{ij}=\max\lbrace 0,u_i(A_j)-u_i(A_i)\rbrace$ be the envy of agent $i$ of agent $j$ and $e(A)=\max\lbrace e_{ij}|i,j\in N\rbrace$ be the maximum envy. The \emph{Lipton-- algorithm} proceeds in $m$ rounds. In the first round, consider bad $o_1$. We allocate $o_1$ to some agent arbitrarily, obtaining $A_1$.  Clearly, the maximum envy in $A_1$ is at most $\alpha$ Suppose at the end of round $(j-1)$, the bads $o_1$ to $o_{j-1}$ have been allocated to agents, obtaining $A_{j-1}$ such that $e(A_{j-1})\leq \alpha$. At round $j$, the algorithm calls a sub-routine with input $A_{j-1}$ and output $A_j$. We have that $e(A_j)\leq e(A_{j-1})\leq\alpha$ and there is an agent, say 1, such that agent 1 is envy free in $A_j$. We then allocate bad $o_j$ to agent 1 in $A_j$. Let $B_j=(A_{j1}\cup\lbrace o_{j}\rbrace,A_{j2}\ldots,A_{jn})$. We next show that the maximum envy in $B_j$ is bounded by $\alpha$. For any two agents $k,h\in N\setminus\lbrace 1\rbrace$, we have that $e_{kh}(B_j)=e_{kh}(A_j)\leq e(A_j)\leq\alpha$. For any agent $k\in N\setminus\lbrace 1\rbrace$, we have that $e_{k1}(B_j)=\max\lbrace 0,u_k(A_{j1}\cup\lbrace o_j\rbrace)-u_k(A_{jk})\rbrace\leq\max\lbrace 0,u_k(A_{j1})-u_k(A_{jk})\rbrace=e_{k1}(A_j)\leq\alpha$ as the preferences are anti-monotone. And, since $e_{1k}(A_j)=0$, we have that $e_{1k}(B_j)=\max\lbrace 0,u_1(A_{jk})-u_i(A_{j1}\cup\lbrace o_j\rbrace)\rbrace\leq \max\lbrace 0,\alpha+u_1(A_{jk})-u_1(A_{j1})\rbrace\leq\alpha$ by the choice of $\alpha$. Finally, \emph{Lipton-- algorithm} runs in $O(mn^3)$ time.
\myqed
\end{myproof}

The \emph{Lipton-- algorithm} allocates each bad to an agent who does not envy any other agent. However, such an agent may or may not have envy of some other agent after the bad is allocated to them. In any case, they remain envy free up to some bad of any other agent. Therefore, the returned allocation is 1EF.

\begin{mycorollary}\label{cor:one}
With anti-monotone valuations, an 1EF allocation can be computed in $O(mn^3)$ time.
\end{mycorollary}

We continue with XEF. With general valuations, we post the existence of an XEF allocation as an \emph{open} problem. Instead, we show that such an allocation always exists in two special cases. For this purpose, we receive an inspiration from the \emph{leximin++ solution}. With identical valuations or 2 agents, this solution is an EFX allocation in the negated problem with goods (Theorems 4.2 and 4.3 in \cite{plaut2018}). By Remark~\ref{rem:one}, with identical valuations, it is therefore an XEF allocation in the initial problem with bads. By Remark~\ref{rem:two}, with 2 agents but possibly distinct valuations, this solution may not be an XEF allocation in this problem. For this reason, we propose a \emph{new} solution that we can directly apply to a given problem with bads. Our \emph{leximax--} solution is such a solution. 

These two solutions relate in various ways. First of all, the \emph{leximin++ solution} uses a total ordering $\prec_{++}$ between allocations in which agents are ordered by increasing valuations (ties are broken arbitrarily but consistently). By comparison, the \emph{leximax-- solution} uses a total ordering $\prec_{--}$ between allocations ordered by decreasing valuations (ties are broken arbitrarily but consistently). We next define these operators.

\emph{$\prec_{++}$ operator ($\prec_{--}$ operator)}: We have $A\prec_{++} B$ ($A\prec_{--} B$) iff, there is $l\in[1,n]$, the $k$th valuation in $A$ equals the $k$th valuation in $B$ and the size of the $k$th bundle in $A$ equals the size of the $k$th bundle in $B$ for each $k<l$, and the $l$th valuation in $A$ is smaller than the $l$th valuation in $B$ or the size of the $l$th bundle in $A$ is smaller (larger) than the size of the $l$th bundle in $B$, with valuation being checked before bundle size. 

On the plus side, even with distinct valuations, an allocation in the problem with bads is greater wrt $\prec_{--}$ than another allocation iff the latter one in the negated problem with goods is greater wrt $\prec_{++}$ than the former one. We prove this in Lemma~\ref{lem:one}. 

\begin{mylemma}\label{lem:one}
Let $A,B$ be two allocations in a problem. We have that $A\prec_{--} B$ in a problem iff $B\prec_{++} A$ in the negated problem.
\end{mylemma}

\begin{myproof}
Let $A$ be an allocation of items. In the problem with bads, let $u_{i_1}(A_{i_1})\geq\ldots\geq u_{i_n}(A_{i_n})$ and $u_{j_1}(B_{j_1})\geq\ldots\geq u_{j_n}(B_{j_n})$ be the decreasing valuation orders in $A$ and $B$ respectively. We observe that, in the negated problem with goods, these orders are reversed and increasing simply because $-u_{i_1}(A_{i_1})\leq\ldots\leq -u_{i_n}(A_{i_n})$ and $-u_{j_1}(B_{j_1})\leq\ldots\leq -u_{j_n}(B_{j_n})$ hold. Suppose now that $A\prec_{--} B$ holds in the problem with bads and consider $l\in[1,n]$ such that $u_{i_k}(A_{i_k})=u_{j_k}(B_{j_k})$ and $|A_{i_k}|=|B_{j_k}|$ for each $k<l$, and $u_{i_l}(A_{i_l})<u_{j_l}(B_{j_l})$, or $u_{i_l}(A_{i_l})=u_{j_l}(B_{j_l})$ and $|A_{i_l}|>|B_{j_l}|$. In the negated problem with goods, for each $k<l$, we have $-u_{i_k}(A_{i_k})=-u_{j_k}(B_{j_k})$ and $|A_{i_k}|=|B_{j_k}|$. For $l$, we have $-u_{i_l}(A_{i_l})>-u_{j_l}(B_{j_l})$, or $-u_{i_l}(A_{i_l})=-u_{j_l}(B_{j_l})$ and $|A_{i_l}|>|B_{j_l}|$. Consequently, $B\prec_{++} A$ holds in the negated problem with goods. In a similar manner, we can prove that if $B\prec_{++} A$ holds, then $A\prec_{--} B$ holds.
\myqed
\end{myproof}

By Lemma~\ref{lem:one}, there is a 1-to-1 correspondence between the minimum allocations wrt $\prec_{--}$ in a problem with bads (i.e.\ the \emph{leximax-- solutions}) and the maximum allocations wrt $\prec_{++}$ in the negated problem with goods (i.e.\ the \emph{leximin++ solutions}). Thus, with identical valuations or 2 agents and possibly distinct valuations, the \emph{leximax-- solution} is also an XEF allocation. We next summarize these corollaries.

\begin{mycorollary}\label{cor:two}
With anti-monotone identical valuations, the \emph{leximax-- solution} is an XEF allocation.
\end{mycorollary}

\begin{mycorollary}\label{cor:three}
With 2 agents and anti-monotone valuations, the \emph{leximax-- solution} is an XEF allocation.
\end{mycorollary}

On the minus side, an EFX allocation may not be a \emph{leximin++ solution} in a problem with 2 agents and identical valuations. To see this, consider the problem in Example 4.3 in \cite{barman2018}. Even more, this solution may not be EFX with 3 agents and additive distinct valuations (see Figure 5 in \cite{plaut2018}). Similarly, an XEF allocation may not be a \emph{leximax-- solution}. To see this, take the problem in Example 4.3 in \cite{barman2018} and negate it. Also, this solution does not guarantee fairness with distinct valuations. 

Finally, we look at the 0/-1 additive case. Similarly, as for an EFX allocation with 0/1 additive valuations for goods in Proposition~\ref{pro:one}, an XEF allocation in this setting can be computed in linear time. For this purpose, we can run the \emph{responsive draft algorithm} on the reduced problem with bads.

\begin{myproposition}\label{pro:three}
With 0/-1 additive valuations, an XEF allocation can be computed in $O(m)$ time.
\end{myproposition}

\subsection{1EF, XEF and Efficiency} We start again with additive valuations. For similar reasons as for goods, we focus only on the Nash welfares and the egalitarian welfares. We start with the egalitarian welfare and egalitarian diswelfare. For the egalitarian welfare, an mEW allocation gives all bads to a single agent and thus cannot be fair. In contrast, an allocation that maximizes the egalitarian welfare might be different to an allocation that maximizes the egalitarian diswelfare. To see this, consider 2 agents, 2 bads with valuations $-1$, $-2$ and $-2$, $-1$. The MEW allocation gives to each agent their most valued bad. This is fair. By comparison, the MEDW allocation gives to each agent their least valued bad. This is not fair or PE. Surprisingly, with identical valuations, both allocations coincide and are thus fair. That is, an MEW allocation is an MEDW allocation and an MEDW allocation is an MEW allocation. To see this, consider an allocation $A$ maximizes the egalitarian diswelfare. This is the same as minimizing the maximum agent's valuation in an allocation. Let us next consider an arbitrary other allocation $B$. We have that $\max_{i\in N} u(B_i)\geq \max_{i\in N} u(A_i)$ holds. Moreover, the total sums of valuations in the allocations $A$ and $B$ are the same because the valuations are identical. Consequently, $\min_{i\in N} u(B_i)\leq \min_{i\in N} u(A_i)$. Hence, $A$ maximizes the egalitarian welfare because of the arbitrary choice of $B$. 

\begin{mycorollary}\label{cor:four}
With additive identical valuations, an MEDW allocation is an MEW allocation and an MEW allocation is an MEDW allocation.
\end{mycorollary}

For the egalitarian diswelfare, an allocation in a problem with bads minimizes (maximizes) this welfare iff it minimizes (maximizes) the egalitarian welfare in the negated problem with goods. Thus, an mEDW allocation gives all bads to a single agent and thus is not fair. In contrast, by Lemma~\ref{lem:one}, the \emph{leximax--} solution minimizes the maximum agent's valuation in an allocation. That is, it maximizes the minimum agent's disvaluation in an allocation. Hence, by Corollary~\ref{cor:two}, an MEDW allocation could be XEF, PE and MUW with identical valuations. 

\begin{mycorollary}\label{cor:five}
With $m$ bads and additive identical valuations, an XEF, PE, MUW and MEDW allocation can be computed in $O(2^m)$ time.
\end{mycorollary}

We proceed with the Nash welfare and Nash diswelfare. For the Nash welfare, we conclude two simple negative results. An allocation that minimizes the Nash welfare may give all bads to a single agent and thus not be fair. Unlike for goods in Theorem~\ref{thm:one}, an MNW allocation with bads may not be XEF with identical valuations. To see this, consider 3 agents, 4 bads and valuations $-2$, $-1$, $-1$ and $-1$. An MNW allocation gives bads to 2 agents only and at least one of these agents receives 2 bads. This is not XEF. 

For the Nash diswelfare, an allocation in a problem with bads minimizes (maximizes) this welfare iff it minimizes (maximizes) the Nash welfare in the negated problem with goods. An MNW allocation is EF1 and PE in the latter problem. By comparion, an MNDW allocation may not be 1EF or PE in the former problem. To see this for PE, consider 2 agents and 2 bads with valuations $-1$, $0$ and $0$, $-1$. The MNDW allocation gives to each agent their least valued bad. This allocation can be Pareto improved by swapping the bads. To see this for 1EF, consider 2 agents and 3 bads with valuations $-1$, $0$, $0$ and $0$, $-1$, $-1$. The MNDW allocation gives to each agent their least valued bads. This allocation is not 1EF as agent 2 envies agent 1 even after removing a single bad from their bundle. Interestingly, with identical valuations, we can restore fairness and efficiency. Corollary~\ref{cor:six} follows by Theorem~\ref{thm:one} and Remark~\ref{rem:one}.

\begin{mycorollary}\label{cor:six}
With additive identical valuations, an MNDW allocation is XEF, PE and MUW.
\end{mycorollary}

Further, we observe that an mNDW allocation may violate 1EF (or XEF) even with identical valuations. To see this, conside 2 agents and 4 identical bads. The mNDW allocation gives 1 bad to one of the agents and 3 bads to the other agent. But then the agent with 3 bads is not 1EF of the other agent. Interestingly, such an allocation is guarantees to be PE even with any number of agents and possibly distinct valuations.

\begin{mytheorem}\label{thm:six}
With additive valuations, an mNDW allocation is PE.
\end{mytheorem}

\begin{myproof}
Let $A$ be an mNDW allocation in the initial problem. We observe that $A$ gives each bad valued with zero by some agents to such an agent. Therefore, the minimum value of the Nash diswelfare of a given problem with bads is equal to the minimum value of the Nash diswelfare of the reduced problem with bads. If we suppose that there is another allocation in the initial problem that Pareto dominates $A$, then we reach a contradiction with the minimality of the diswelfare under $A$ in the reduced problem. Hence, $A$ is PE in the reduced problem and in the initial problem.\myqed
\end{myproof}

We next discuss a few computational questions. Computing MNDW allocations is intractable even with 2 agents, $m>n$ bads and identical valuations \cite{nguyen2013,ramezani2009}. By comparison, computing mNDW allocations is tractable in general. To see this, we can allocate all bads to a single agent. The value of the Nash diswelfare in this allocation is 0. Unfortunately, this allocation is very unfair. In response, we may first want to allocate each bad valued with zero by some agents to such an agent and then compute an mNDW allocation in the reduced problem over a largest set of agents to which one can simultaneously provide a negative valuation for a good (i.e.\ an mNDW$^{-}$ allocation). This approach is not computational fast. And, it may also not deliver any fairness guarantees. To see this, consider again the problem with 2 agents and 4 identical bads. In contrast, with $m\leq n$ bads, this 
approach is computationally feasible. Moreover, we gain fairness in addition in this setting.

\begin{mytheorem}\label{thm:seven}
With $m\leq n$ bads and additive valuations, an XEF, PE and mNDW$^{-}$ allocation can be computed in $O(n^3)$ time.
\end{mytheorem}

\begin{myproof}
Let us consider the reduced problem. WLOG, let there be $n$ goods in it. Further, negate it and apply $\ln$-function to each valuation in the negated problem. Consider the weighted, undirected and complete bipartite graph that corresponds to this problem with logarithmic valuations. 
A minimum weight perfect matching in this graph corresponds to an mNDW allocation in the reduced problem over a largest set of agents with negative valuations for goods. We can compute such a matching in $O(n^3)$ time by using the Hungarian method. PE of this allocation follows by Theorem~\ref{thm:six}. Finally, XEF of it follows as each agent gets at most one bad in it. 
\myqed
\end{myproof}

``Exact'' Nash efficiency gives us no fairness guarantees unless the valuations are identical. Moreover, it may take long time to compute an allocation that is ``exact'' Nash efficient. As for goods, we might be satisfied with an allocation that is ``almost'' Nash efficient and fair instead. We show that the \emph{Alg-Identical algorithm} returns such an allocation. If we have access to the valuations of the agents for bads, then we view this algorithm as an alternative to the \emph{Alg-Identical algorithm}. If each bad arrives in an online manner with no information about the bads to arrive in future, then we might run the \emph{Alg-Identical-- algorithm} directly to the problem with bads. In either case, these two algorithms relate in an interesting way. 

\begin{mylemma}\label{lem:two}
Let $A_1$ be the allocation returned by \emph{Alg-Identical-- algorithm} in a problem with bads. Let $A_2$ be the allocation returned by \emph{Alg-Identical algorithm} in the negated problem with goods. With identical valuations, we have $A_{1i}=A_{2i}$ for each $i\in N$.
\end{mylemma}

\begin{myproof}
Let us suppose that both algorithms use $(1,\ldots,n)$ as a tie-breaker. First, let us consider a problem with bads and its negated version with goods. As the valuations are identical, both algorithms use the same ordering $o$ of items. Suppose that they allocate the first $(j-1)$ items in this ordering to the same agents. Further, let $A_{(j-1)}$ be this allocation. It suffices to show that the $j$th item in $o$ in these problems is allocated to the same agent by both algorithms. Let \emph{Alg-Identical-- algorithm} allocates the $j$th bad in $o$ to agent $i$ in the problem with bads. Hence, $i=\max_{k\in N} u_k(A_{(j-1)})$. In the negated problem with goods, we have that $i=\min_{k\in N} -u_k(A_{(j-1)})$. We conclude that \emph{Alg-Identical algorithm} allocates the $j$th good in $o$ to agent $i$ in the negated problem with goods. The result follows.
\myqed
\end{myproof}

By Lemma~\ref{lem:two}, we conclude that the \emph{Alg-Identical-- algorithm} returns an ``almost'' MNDW allocation up to a tight factor of $1.061$ in a given problem with bads as the \emph{Alg-Identical algorithm} returns an ``almost'' MNW allocation up to this factor in the negated problem with goods. By Remark~\ref{rem:one}, such an allocation satisfies XEF. Moreover, it is trivially PE and MUW. As for goods in Theorem~\ref{thm:three}, with $m\leq n$, we additionally gain MEW as the algorithm gives to each agent at most one bad.

\begin{mycorollary}\label{cor:seven}
With $m>n$ bads and additive identical valuations, an XEF, PE, MUW and ``almost'' MNDW allocation can be computed in $O(m)$ time.
\end{mycorollary}

\begin{mycorollary}\label{cor:eight}
With $m\leq n$ bads and additive identical valuations, an XEF, PE, MUW, MEW and ``almost'' MNDW allocation can be computed in $O(m)$ time.
\end{mycorollary}

By Corollaries~\ref{cor:five},~\ref{cor:seven} and~\ref{cor:eight}, we observe a computational trade-off between MEW and ``almost'' MNDW for bads. By Observations~\ref{obs:one},~\ref{obs:two} and Theorem~\ref{thm:three}, there is a similar trade-off between MEW and ``almost MNW'' for goods.
As for goods, we might be interested in achieving only 1EF and PE in the absence of MEW or MNDW. With any number of agents and identical valuations, we can compute such allocations in linear time by Corollaries~\ref{cor:seven} and~\ref{cor:eight}. With 2 agents and possibly distinct valuations, we can do this with the algorithm from Theorem 2 in \cite{aziz2018ai3}. With 3 or more agents, we conjecture that there is \emph{no} such polynomial algorithm (unless $\P=\NP$). However, with 0/-1 valuations, we can compute an XEF, PE, MUW, MEW and mNDW allocation in linear time with the \emph{responsive draft algorithm} on the reduced problem. This result follows by Proposition~\ref{pro:three}.

We next consider anti-monotone valuations. Unlike for additive valuations in Corollaries~\ref{cor:five} and~\ref{cor:six}, we might not be able to achieve both 1EF and PE with \emph{zero} marginal identical valuations (and, therefore, 1EF and MUW or MNW).

\begin{mytheorem}\label{thm:eight}
Even with 2 agents and (zero marginal) anti-monotone identical valuations, there are problems in which \emph{no} 1EF allocation is PE.
\end{mytheorem}

\begin{myproof}
Let us consider $N=\lbrace 1,2\rbrace$ and $O=\lbrace o_1,o_2,o_3\rbrace$. Further, let $u(O)=-5/2$, $u(\lbrace o_1,o_2\rbrace)=u(\lbrace o_2,o_3\rbrace)=u(\lbrace o_1,o_3\rbrace)=-2$, $u(o_1)=u(o_2)=u(o_3)=0$. Consider the allocation $A=(O,\emptyset)$. This allocation and the one in which agents swap their bundles are the only PE allocations. However, the allocation $A$ is not 1EF (and XEF) as $u(A_1\setminus\lbrace o_1\rbrace)=-2<0=u(A_2)$, $u(A_1\setminus\lbrace o_2\rbrace)=-2<0=u(A_2)$ and $u(A_1\setminus\lbrace o_3\rbrace)=-2<0=u(A_2)$. \myqed
\end{myproof}

With \emph{non-zero} marginal valuations, we give similar impossibility results. By Corollary~\ref{cor:six}, 1EF (even XEF) and MNDW can be guaranteed in combination with additive identical valuations. This is not true with anti-monotone identical valuations. To see this, negate the problem in Theorem~\ref{thm:four}. The MNDW allocation in this problem is not 1EF. Finally, by Lemma~\ref{lem:one}, the \emph{leximax-- solution} is an XEF and MEDW allocation with any number of agents and identical valuations or 2 agents and possibly distinct valuations. With 3 agents, an XEF allocation may not be MEDW even with additive valuations (take the problem in Figure 5 in \cite{plaut2018} and negate it). Unlike for additive valuations in Corollary~\ref{cor:four}, an MEW allocation may not be an MEDW allocation with anti-monotone valuations. Moreover, an MEW allocation may not be 1EF.

\begin{mytheorem}\label{thm:nine}
Even with 2 agents and (non-zero marginal) anti-monotone identical valuations, there are problems in which \emph{no} 1EF allocation is MEW.
\end{mytheorem}

\begin{myproof}
Let us consider $N=\lbrace 1,2\rbrace$ and $O=\lbrace o_1,o_2,o_3,$ $o_4\rbrace$. Further, let $u(O)=-5$, $u(\lbrace o_1,o_2,o_3\rbrace)=-3$, $u(\lbrace o_1,o_2,o_4\rbrace)=u(\lbrace o_1,o_3,o_4\rbrace)$ $=u(\lbrace o_2,o_3,o_4\rbrace)=-4$, $u(\lbrace o_1,o_2\rbrace)=u(\lbrace o_1,o_3\rbrace)=u(\lbrace o_2,o_3\rbrace)=-2$, $u(\lbrace o_1,o_4\rbrace)$ $=u(\lbrace o_2,o_4\rbrace)=u(\lbrace o_3,o_4\rbrace)=-7/2$, $u(o_1)=u(o_2)=u(o_3)=-3/2$ and $u(o_4)=-1$. Consider the allocation $A=(\lbrace o_1,o_2,o_3\rbrace,\lbrace o_4\rbrace)$. This allocation and the one in which agents swap their bundles are the only two allocations that maximize the egalitarian welfare of $-3$. However, the allocation $A$ is not 1EF (and XEF) as $u(A_1\setminus\lbrace o_1\rbrace)=-2<-1=u(A_2)$, $u(A_1\setminus\lbrace o_2\rbrace)=-2<-1=u(A_2)$ and $u(A_1\setminus\lbrace o_3\rbrace)=-2<-1=u(A_2)$. Finally, note that the allocation $A$ does not maximize the egalitarian diswelfare. Such an allocation gives two bads to each agent and is XEF. 
\myqed
\end{myproof}

\section{Conclusions}

We next summarize our work. We \emph{generalized} existing notions of ``almost'' envy freeness for fair division with goods (i.e.\ EF1 and EFX). We also proposed similar notions for fair division with bads (i.e.\ 1EF and XEF). For fair division with goods, an MNW allocation is EF1 and PE. However, it is intractable even with identical valuations. We showed that it is even EFX in this setting. We further presented several tractable cases in which almost fairness and efficiency can be achieved in combination (e.g.\ $m\leq n$ goods, 0/1 additive valuations). For fair division with bads, we showed that most existing algorithms for goods return unfair allocations. We thus proposed \emph{new} algorithms for this setting, e.g.\ \emph{Lipton-- algorithm}, \emph{leximax-- solution} and \emph{Alg-Identical-- algorithm}. For example, the \emph{Lipton-- algorithm} returns an 1EF allocation with arbitrary anti-monotone valuations. Moreover, this allocation has bounded maximum envy. We also study the egalitarian diswelfare and Nash diswelfare in this setting. Like for goods and MNW, an allocation that maximizes the Nash diswelfare is XEF and PE with identical valuations. Unlike for goods and MNW, it may be neither 1EF nor PE with distinct valuations. Interestingly, in this case, an allocation that minimizes this welfare is guaranteed to be PE. As for goods, we reported several tractable cases (e.g.\ $m\leq n$ bads, 0/-1 additive valuations). Finally, we proved a number of \emph{new} impossibility results. For example, EF1 and Nash efficiency cannot be achieved together even with 2 agents and (non-zero marginal) monotone identical valuations.

\newpage

\bibliographystyle{splncs04}
\bibliography{almost}

\newpage

\section{Appendix: remarks}

\begin{myproofremone} With identical valuations, an allocation in a problem with bads is XEF (1EF) iff it is EFX (EF1) in the negated problem with goods. Let $A$ be an EFX allocation in the negated problem with goods. Let $u(A_i)$ denote the valuation of each $k\in N$ for $A_i$ in $A$. For the sake of contradiction, suppose that $A$ is not XEF (1EF) in the initial problem with bads. Let $v(A_i)$ denote the valuation of agent $k\in N$ for $A_i$ in $A$ in the initial problem with bads. Note that $v(A_i)=-u(A_i)$. Consequently, there are two agents $i,j\in N$ and one item (each item) $o\in A_i$ such that $v(A_i\setminus\lbrace o\rbrace)<v(A_j)$. We derive that $u(A_j)<u(A_i\setminus\lbrace o\rbrace)$. As $A$ is EFX (EF1) in the negated problem with goods, we have $u(A_j)\geq u(A_i\setminus\lbrace o\rbrace)$. This is a contradiction. Therefore, $A$ is XEF (1EF) in the initial problem with bads. Similarly, we can show that if the allocation $A$ is an XEF (1EF) allocation in a problem with bads then the allocation $A$ is EFX (EF1) in the negated problem with goods.\myqed \end{myproofremone}

\begin{myproofremtwo} Even with 2 agents and additive distinct valuations, an XEF allocation in a problem with bads may not be an EFX allocation in the negated problem with goods, and an EFX allocation in the negated problem with goods may not be an XEF allocation in the problem with bads. The same holds for EF1 and 1EF. Let us consider $N=\lbrace 1,2\rbrace$ and $O=\lbrace o_1,o_2,o_3\rbrace$. We consider two settings. First, suppose the items are goods and let agent 1 likes items $o_1$, $o_2$ and $o_3$ with 5, 2 and 3 respectively, whilst agent 2 likes them with 1, 4 and 5. Consider next the allocation $A$ that gives item $o_1$ to agent 1 and items $o_2$ and $o_3$ to agent 2. This allocation is EFX and EF1. However, this allocation is not XEF or even 1EF in the negated problem as agent 2 envies agent 1 even after the removal of any item from their bundle. Second, suppose the items are bads and let agent 1 likes items $o_1$, $o_2$ and $o_3$ with $-5$, $-2$ and $-3$ respectively, whilst agent 2 likes them with $-1$, $-4$ and $-5$. Consider next the allocation that gives item $o_1$ to agent 2 and items $o_2$ and $o_3$ to agent 1. This allocation is XEF and 1EF. However, this allocation is not EFX or even EF1 in the negated problem as agent 1 envies agent 2 after the removal of any item from their bundle. \myqed \end{myproofremtwo}

\newpage

\section{Appendix: propositions}

\begin{myproofproone} With 0/1 additive valuations, an EFX allocation always exists and can be computed in $O(m)$ time. Let us consider the allocation returned by a \emph{responsive draft algorithm}. We claim that this allocation is EFX. To see this, we make two simple observations. First, for each $j\in[1,m]$, $o_j$ is allocated to an agent who likes it with 1. Second, for every two agents $i,j\in N$ with $i\not=j$, either agent $i$ receives at most one more good that agent $j$ likes with 1 \emph{or} agent $j$ receives at most one more good that agent $i$ likes with 1. WLOG, suppose that agent $i$ receives at most one more good that agent $j$ likes with 1. We consider two cases. In the first case, suppose that agent $i$ receives $k$ goods that agent $j$ likes with 1 and agent $j$ receives $(k-1)$ goods that they like with 1. Therefore, agent $i$ is EF and thus EFX of agent $j$. Also, by ``removing'' whichever good that agent $j$ likes with 1 from the bundle of agent $i$, we conclude that agent $j$ assigns a value of $(k-1)$ to the bundle of agent $i$ of remaining goods. But this is the same value as they assign to their own bundle. Hence, agent $j$ is EFX of agent $i$. In the second case, suppose that agent $i$ receives $k$ goods that agent $j$ likes with 1 and agent $j$ receives $k$ goods that they like with 1. We conclude that both agents are EF and thus EFX of each other.\myqed \end{myproofproone}

\begin{myproofprotwo} With 0/1 additive valuations, an EFX, utilitarian and Nash efficient allocation always exists and can be computed in $O(2m(n+1)\cdot\ln(mn))$ time. Let $u_k(A)$ be the valuation of agent $k\in N$ in a given allocation $A$. Further, let us consider the allocation $A$ returned by the \emph{Alg-Binary algorithm}. We next show that $A$ is egalitarian efficient. We first argue that the egalitarian welfare in $A$ is greater than the egalitarian welfare in any allocation $B$ that does not maximize the Nash welfare. For the sake of contradiction, assume that $\min_{k\in N} u_k(B)>\min_{k\in N} u_k(A)$. Let us consider $N^{\prime}=\lbrace i|i\in N,u_i(B)\geq u_i(A)\rbrace$ and $N^{\prime\prime}=\lbrace i|i\in N,u_i(B)<u_i(A)\rbrace$. As the Nash welfare is not optimal under $B$, we conclude $N^{\prime\prime}\not=\emptyset$ and, for each $i\in N^{\prime}$ and $k\in N^{\prime\prime}$, $u_i(B)\geq u_k(B)$. Otherwise, we can increase the Nash welfare in $A$. The algorithm with input $B$ and output $A$ moves goods from agents with more goods in $B$ and $N^{\prime}$ to agents with less goods in $B$ and $N^{\prime\prime}$ until, for each $i\in N^{\prime}$ and $k\in N^{\prime\prime}$, $u_i(A)\geq u_k(A)$ (see \cite{barman2018} for details). Let us further consider agents $i=\arg\min_{k\in N} u_k(B)$ and $j=\arg\min_{k\in N} u_k(A)$. If agent $i\in N^{\prime\prime}$, then agent $j\in N^{\prime\prime}$. Hence, we derive $u_j(A)>u_j(B)\geq u_i(B)$ which contradicts our assumption. Agent $i$ cannot be in $N^{\prime\prime}$. If agent $i\in N^{\prime}$ and agent $j\in N^{\prime\prime}$, then we derive $u_j(A)>u_j(B)\geq u_i(B)$ which again contradicts our assumption. Agent $j$ cannot be in $N^{\prime\prime}$ Therefore, it must be the case that both agents $i,j\in N^{\prime}$. For each $k\in N^{\prime\prime}$, $u_k(A)\geq u_j(A)$ and $u_k(B)\geq u_i(B)$. Moreover, for each $k\in N^{\prime\prime}$, $u_j(A)\geq u_k(A)$ and $u_i(B)\geq u_k(B)$. Therefore, for each $k\in N^{\prime\prime}$, we derive $u_j(A)=u_k(A)$, $u_i(B)=u_k(B)$ and $u_k(B)<u_k(A)$ which contradicts our assumption.

We next argue that the egalitarian welfare in $A$ is the same as the egalitarian welfare in any allocation $C$ that maximizes the Nash welfare. For this reason, we show that the allocation $C$ can be obtained from $A$ by permuting the names of the agents. Let us consider agent $j=\arg\min_{k\in N} u_k(A)$. For each agent $i\in N\setminus\lbrace j\rbrace$, we consider three cases in the allocation $A$. In the first case, let $u_i(A)>u_j(A)+1$. If agent $j$ likes at least one good in $A_i$, we can move such a good from $A_i$ to $A_j$ and thus increase the Nash welfare. Therefore, agent $j$ does not like any good in $A_i$. Furthermore, by moving goods from $A_i$ to $A_j$, the Nash welfare in $A$ decreases and the egalitarian welfare in $A$ cannot increase. In the second case, let $u_i(A)=u_j(A)+1$. Now, agent $j$ may like some goods from $A_i$. If we move one such good from $A_i$ to $A_j$, then the Nash and egalitarian welfares in $A$ remain the same. If we move two or more such goods from $A_i$ to $A_j$, then the Nash and egalitarian welfares in $A$ decrease. In the third case, let $u_i(A)=u_j(A)$. If we move goods from $A_i$ to $A_j$, then the Nash and egalitarian welfares in $A$ decrease. We conclude that each allocation that maximizes the Nash welfare such as $C$ can be obtained from $A$ by permuting the names of the agents. Hence, the egalitarian welfare in $C$ is equal to the egalitarian welfare in $A$. The result follows.
\myqed \end{myproofprotwo}

\begin{myproofprothree} With 0/-1 additive valuations, an XEF allocation always exists and can be computed in $O(m)$ time. We can first allocate each bad valued with zero by some agent to such an agent and then run a \emph{responsive draft algorithm} with the remaining items, say $o_1$ to $o_h$. We claim that the returned allocation is XEF. To see this, we make a simple observation. Within bads $o_{1}$ to $o_h$, for every two agents $i,j\in N$ with $i\not=j$, agent $i$ receives at most one more bad than agent $j$ \emph{or} agent $j$ receives at most one more bad than agent $i$. WLOG, suppose that agent $i$ receives at most one more bad than agent $j$. We consider two cases. In the first case, suppose that agent $i$ receives $k$ bads that they dislike with -1 and agent $j$ receives $(k-1)$ bads that they dislike with -1. Therefore, agent $j$ is EF and thus XEF of agent $i$. Also, by ``removing'' whichever bad that agent $i$ dislikes with -1 from their bundle, we conclude that agent $i$ assigns a value of $-(k-1)$ to their bundle of remaining bads. But this is the same value that they assign to the bundle of agent $j$. Hence, agent $i$ is XEF of agent $j$. In the second case, suppose that agent $i$ receives $k$ bads that they dislike with -1 and agent $j$ receives $k$ bads that they dislike with -1. We conclude that both agents are EF and thus XEF of each other.\myqed \end{myproofprothree}

\end{document}